\documentclass[prl,twocolumn]{revtex4}
\usepackage{graphicx}
\usepackage{amsmath}
\usepackage{dcolumn}
\usepackage{color}
\usepackage{epstopdf}

\def\v{\textbf{v}}

\def\B{\textbf{B}}

\def\e{\textbf{e}}
\def\pa{\partial}

\def\r{\textbf{r}}

\begin{document}

\title{Fast dynamos in spherical boundary-driven flows}

\author{I. V. Khalzov}
\affiliation{Center for Magnetic Self Organization in Laboratory and Astrophysical Plasmas, University of Wisconsin-Madison, 1150 University Avenue, Madison, Wisconsin 53706, USA}

\author{C. M. Cooper}
\affiliation{Center for Magnetic Self Organization in Laboratory and Astrophysical Plasmas, University of Wisconsin-Madison, 1150 University Avenue, Madison, Wisconsin 53706, USA}

\author{C. B. Forest}
\affiliation{Center for Magnetic Self Organization in Laboratory and Astrophysical Plasmas, University of Wisconsin-Madison, 1150 University Avenue, Madison, Wisconsin 53706, USA}

\date{\today}

\begin{abstract}
We numerically demonstrate  the feasibility of  kinematic fast dynamos for a class of time-periodic axisymmetric flows of conducting fluid confined inside a sphere. The novelty of our work is in considering the realistic flows, which are self-consistently determined from the Navier-Stokes equation with specified boundary driving. Such flows can be achieved in a new  plasma  experiment, whose spherical boundary is capable of differential driving of plasma flows in the azimuthal direction. We show that magnetic fields are self-excited over a range of flow parameters such as amplitude and frequency of flow oscillations, fluid Reynolds ($Re$) and magnetic  Reynolds ($Rm$) numbers. In the limit of large $Rm$, the growth rates of the excited magnetic fields are of the order of the advective time scales and practically independent of $Rm$, which is an indication of the fast dynamo.
\end{abstract}

\maketitle

It is now widely accepted that the observed magnetic fields of various astrophysical systems are due to the dynamo mechanism associated with the flows of highly conducting fluids in their interiors \cite{Moffatt_1978, Rudiger_2004}.  In most of these  systems  the dynamos appear to be \emph{fast}, i.e., they act on the fast advective time scales of the underlying flows, rather than on the slow resistive or intermediate time scales \cite{Vainshtein_1972}. The astrophysical relevance of fast dynamos stimulated intensive theoretical and numerical studies (Refs.~\cite{Childress_1995, Galloway_2003} and references therein), however, they have never been studied experimentally. In this work we present a numerical analysis of fast dynamos in a class of  spherical boundary-driven flows that can be obtained in a real plasma  experiment. 

Several conditions are required for a fast dynamo action. First, the fast dynamos only operate at high values of magnetic Reynolds number $Rm$ (usually, $Rm\gtrsim10^3$), which is the ratio of the resistive and advective time scales. Second, the  flow must be chaotic in order to yield a fast dynamo \cite{Finn_1988, Klapper_1995}. Both these conditions are satisfied in astrophysical flows, most of which are turbulent with extremely high $Rm$. Achieving flows with high $Rm$ is a major obstacle  for experimental demonstration of fast dynamos, since this requires a highly conducting, quickly  flowing fluid (for reference, the maximum  value of $Rm$ in existing liquid metal experiments is  $Rm\sim10^2$). 

Models with prescribed flows exhibiting fast dynamos in the kinematic (linear) stage have been considered in the literature for plane-periodic \cite{Galloway_1992, Hughes_1996, Dorch_2000, Tobias_2008} and spherical shell \cite{Hollerbach_1995, Hollerbach_1998, Livermore_2007, Richardson_2012} geometries. The common feature of the dynamo fields  emerging  in these models  is that  at high $Rm$ their structure is dominated by the rapidly varying small-scale fluctuations. The relationship between the small-scale dynamo and large-scale generation of flux is perhaps the most perplexing issue in dynamo theory. As shown recently in Refs.~\cite{Hughes_2009,Tobias_2013}, an organized large-scale magnetic field can be produced in dynamo simulations if a global shear is also included into a helical plane-periodic flow. 

More realistic models of dynamos include flows which are  self-consistently determined from the Navier-Stokes equation with the appropriate forcing terms \cite{Brummel_1998, Cameron_2006, Livermore_2010}.  In this case, the properties of fast dynamos (in both the linear and nonlinear stages) strongly depend on the magnetic Prandtl number $Pm=\nu/\eta$,  the ratio of kinematic viscosity $\nu$ to resistivity $\eta$ of the medium. In astrophysical applications of fast dynamo theory two different limits are recognized: $Pm\ll1$ (solar convective zone, protoplanetary accretion disks) and $Pm\gg1$ (solar corona, interstellar medium, galaxies and galaxy clusters). In the former limit, the fluid is essentially inviscid on all magnetic scales, and the resulting dynamo is induced by the ``rough" (highly fluctuating) velocity field \cite{Tobias_2012, Schekochihin_2007}. In the latter limit, the small velocity scales are dissipated more efficiently than the small magnetic  scales, and the dynamo develops on the  ``smooth'' (large scale) velocity field \cite{Tobias_2012, Schekochihin_2004}. The latter  circumstance is important in our study: the case of $Pm\ll1$ cannot be addressed  without extensive numerical simulations of MHD turbulence, but the case of $Pm\gg1$ can be modeled by a simple laminar, yet chaotic flow producing a  fast dynamo structure. 

Our present study is motivated by the successful initial operation  of the Madison plasma dynamo experiment (MPDX), which is  designed to investigate dynamos in controllable flows of hot, unmagnetized plasmas \cite{Spence_2009}. In the MPDX, plasma is confined inside a 3 m spherical vessel by an edge-localized multicusp magnetic field. A unique  mechanism for driving plasma flows in the MPDX allows an arbitrary azimuthal velocity  to be imposed along  the boundary  \cite{Collins_2012}. The most advantageous property of plasma is the ability  to vary its characteristics by adjusting experimental controls. Indeed, the confinement properties of this device indicate that the low density, high temperature plasmas can be obtained with values of  $Rm\sim10^4$ and $Pm\sim10^2$ required for a  fast dynamo in a laminar flow. 

The scope of our study is to numerically demonstrate kinematic fast  dynamos for a class of  time-periodic, boundary-driven, axisymmetric flows found self-consistently from the Navier-Stokes equation.  In a sense, these flows are an adaptation of the plane-periodic Galloway-Proctor flow \cite{Galloway_1992} to the geometry of a full sphere with a viable driving force. Several models of  flows resulting in fast dynamos have been considered  in the spherical shells \cite{Hollerbach_1995, Hollerbach_1998, Livermore_2007, Richardson_2012, Livermore_2010}, but analogous studies for a full sphere are lacking. Furthermore, in all previous  studies the flows are either prescribed or driven by artificial forces, so they are non physical and not reproducible in a real experiment.  The class of  flows considered in our study is more practical, since it can be realized in the MPDX by applying the corresponding plasma driving at the edge.  

As a framework, we use incompressible magnetohydrodynamics (MHD), whose dimensionless equations  are  
\begin{eqnarray}
\label{mhd_NS} \frac{\pa\v}{\pa t}&=&\frac{1}{Re}\nabla^2\v-(\v\cdot\nabla)\v-\nabla p,\\
\label{mhd_v} \frac{\pa\tilde\v}{\pa t}&=&\frac{1}{Re}\nabla^2\tilde\v-(\tilde\v\cdot\nabla)\v-(\v\cdot\nabla)\tilde\v-\nabla\tilde p,\\
\label{mhd_b} \frac{\pa\B}{\pa t}&=&\frac{1}{Rm}\nabla^2\B+\nabla\times(\v\times\B),\\
\label{mhd_div} 0&=&\nabla\cdot\v=\nabla\cdot\tilde\v=\nabla\cdot\B.
\end{eqnarray}
Normalized quantities are defined using the radius of the sphere $R_0$ as a unit of length, the typical driving velocity at the edge $V_0$ as a unit of velocity, the turnover time $R_0/V_0$ as a unit of time, the plasma mass density  $\rho_0$, the kinematic viscosity $\nu$ and the magnetic diffusivity $\eta$ (all three assumed to be constant and uniform). Normalization of the magnetic field $\B$ is arbitrary since it enters Eq.~(\ref{mhd_b}) linearly. The fluid Reynolds number $Re$ and the magnetic Reynolds number  $Rm$ are introduced as $Re=R_0V_0/\nu$, $Rm=R_0V_0/\eta$, and their ratio determines magnetic Prandtl number $Pm=Rm/Re$. In this study we consider flows in a range of $Re=100-500$, $Rm=10^3-10^5$ and  $Pm=2-500$, which overlaps with MPDX operating regimes. In these regimes,  MPDX plasma is expected to have relatively small Mach numbers $M=V_0/C_s\approx0.1-0.3$, where $C_s$ is the ion sound speed. This justifies the use of the incompressible model.

\begin{figure}[tbp]
\centering
\includegraphics[scale=1]{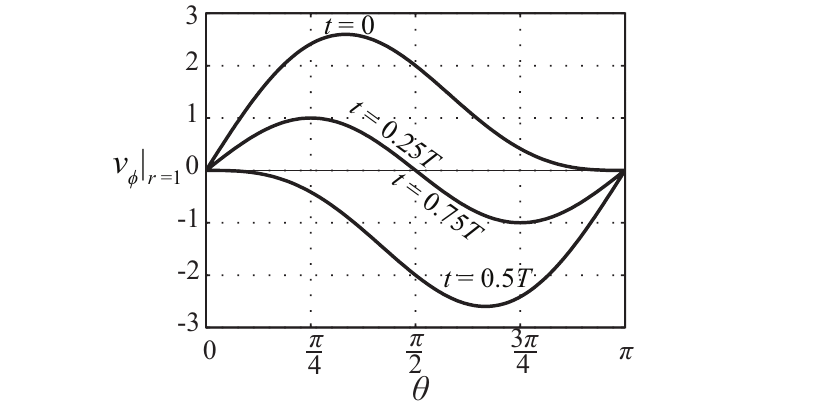}
\caption{Time-periodic boundary-driving azimuthal velocity $v_\phi$ from Eq.~(\ref{BCv}) with oscillation amplitude $a=2$  as a function of polar angle $\theta$  at several moments  over a period $T=2\pi/\omega$. }\label{vel_BC}
\end{figure}

Eqs.~(\ref{mhd_NS})-(\ref{mhd_b}) require appropriate boundary conditions. Eq.~(\ref{mhd_NS}) is the Navier-Stokes equation and we use it to find equilibrium flows $\v$. Since we are interested in the kinematic stage of a dynamo, we do not include the Lorentz force from the dynamo field in Eq.~(\ref{mhd_NS}). We consider only axisymmetric, time-periodic equilibrium flows driven by the following  boundary condition (Fig.~\ref{vel_BC}): 
\begin{equation}\label{BCv}
\v\big|_{r=1}=(\sin2\theta+a\sin\theta\,\cos\omega t)\e_\phi,~~~0\leq\theta\leq\pi,
\end{equation}
in a spherical system of coordinates $(r,\theta,\phi)$. As shown below, for some values of oscillation amplitude $a$ and frequency $\omega$ the corresponding equilibrium flows  are hydrodynamically stable and can lead to fast dynamos. Eq.~(\ref{mhd_v}) is the linearized Navier-Stokes equation for the flow perturbations $\tilde\v$. It is used to establish the hydrodynamic stability properties of the equilibrium flows $\v$. Assuming impenetrable, no-slip wall, we have $\tilde\v\big|_{r=1}=0$. Eq.~(\ref{mhd_b}) is the kinematic dynamo problem with a velocity field $\v$ satisfying Eqs.~(\ref{mhd_NS}), (\ref{BCv}). The choice of magnetic boundary conditions can strongly affect the dynamo action \cite{Khalzov_2012b}. In our present study we assume the non-ferritic, perfectly conducting wall, coated inside with a thin insulating layer. This model reflects the actual boundary in the MPDX. Corresponding conditions for $\B$ are  $B_r\big|_{r=1}=(\nabla\times\B)_r\big|_{r=1}=0$, i.e., normal components of  both  magnetic field and current are zero at the boundary. 

We briefly describe the numerical methods used for solving Eqs.~(\ref{mhd_NS})-(\ref{mhd_b}). The divergence-free fields $\v$, $\tilde{\v}$ and $\B$ are expanded  in a spherical harmonic basis \cite{Bullard_1954} (the resulting equations can be found, for example, in Ref.~\cite{Khalzov_2012c}). The radial discretization of these equations is based on a Galerkin scheme with Chebyshev polynomials \cite{Livermore_2005}. This discretization provides fast convergence: truncation at $N_p=100$ spherical harmonics and $N_r=100$ radial polynomials gives 4 converged significant digits in the dynamo growth rate for  $Rm=50\,000$. To solve the discretized Eq.~(\ref{mhd_NS}) for a time-periodic equilibrium velocity  $\v$ we apply a Fourier time transform to $\v$ and  use the iterative scheme analogous to the one from Ref.~\cite{Khalzov_2012c}. Since the equilibrium velocity is axisymmetric, modes with different azimuthal numbers $m$ are decoupled  in Eqs.~(\ref{mhd_v}), (\ref{mhd_b}). For a time-periodic flow $\v$, these equations constitute the standard Floquet eigenvalue problems \cite{Kuchment_1993}. Eq.~(\ref{mhd_b}) has a solution of the form $\B(\r,t)=e^{\gamma t}\tilde\B(\r,t)$, where $\gamma$ is the complex eigenvalue (Floquet exponent) and $\tilde\B(\r,t)$ is time-periodic with a period $T=2\pi/\omega$ of the flow (solution to Eq.~(\ref{mhd_v}) is analogous).  We solve  Eqs.~(\ref{mhd_v}), (\ref{mhd_b}) in  discretized form using the Arnoldi iteration method, which finds the eigenvalues with the largest real part (growth rate). More details on application of this method to a kinematic  dynamo problem are given in Ref.~\cite{Willis_2004}.  

\begin{figure}[tbp]
\centering
\includegraphics[scale=1]{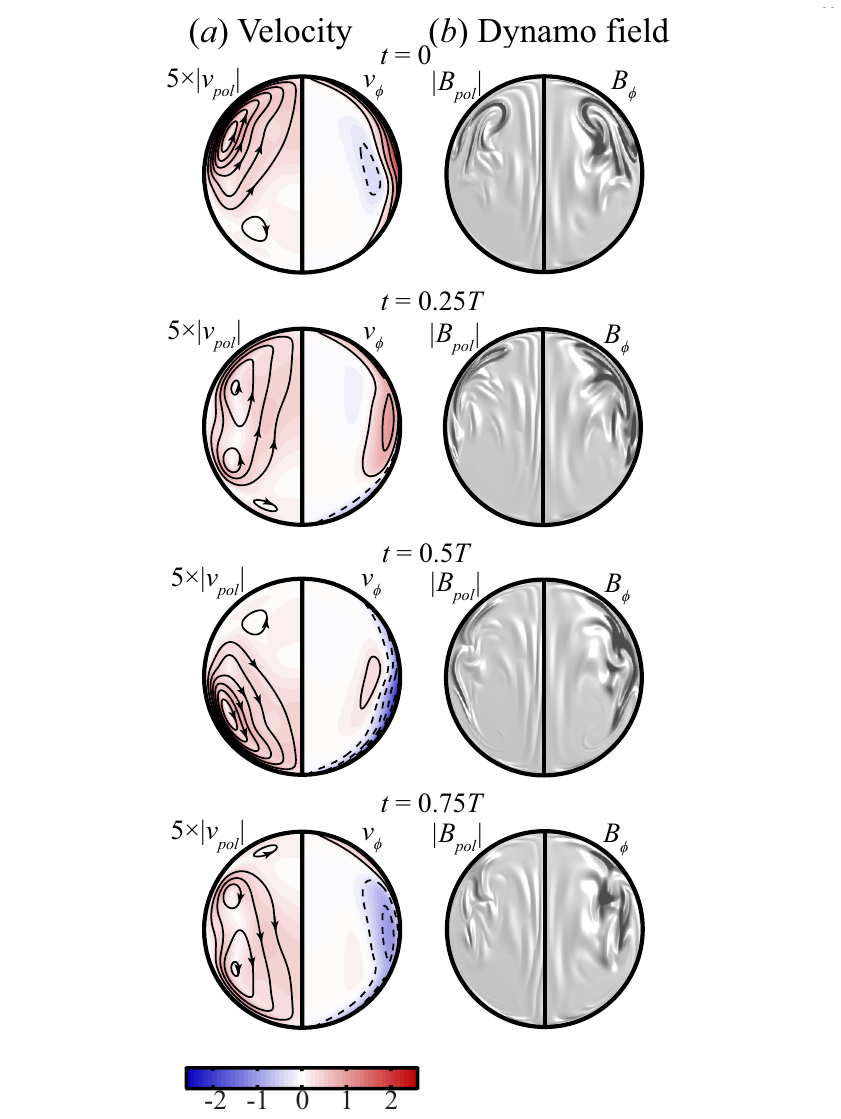}
\caption{Structure of (a) equilibrium velocity  $\v$ and (b) fastest dynamo eigenmode $\B$ (with azimuthal number $m=1$) at four consecutive moments over a  period $T=2\pi/\omega$. The flow parameters are  $a=2$, $\omega=0.6$, $Re=300$, $Rm=30\,000$, the respective  dynamo eigenvalue is $\gamma=0.059627+0.200510i$. Exponential growth and rotation associated with $\gamma$ are removed from the images of $\B$ by a suitable normalization (the resulting $\B$ is periodic with period $T$).  The left half of each figure shows the modulus of the poloidal (lying in meridional plane) component, the right half shows a level plot of the azimuthal component. In velocity figures,  contours with arrows are stream lines of poloidal velocity $v_{pol}$, dashed curves denote levels  of $v_\phi<0$. In dynamo figures, lighting is used to emphasize the small-scale structure of  dynamo field. The listed parameters correspond to the MPDX fully ionized hydrogen plasma with the density $n_0\approx2\cdot10^{12}$~cm$^{-3}$, the ion temperature $T_i\approx1$~eV, the electron temperature $T_e\approx100$~eV, the typical driving velocity  $V_0\approx12.5$~km/s, the driving frequency $\omega/2\pi\approx0.8$~kHz and the dynamo growth rate $\gamma_r\approx500$~s$^{-1}$  (see Ref.~\cite{Spence_2009} for details).}\label{tot_new2}
\end{figure}  

\begin{figure}[tbp]
\centering
\includegraphics[scale=1]{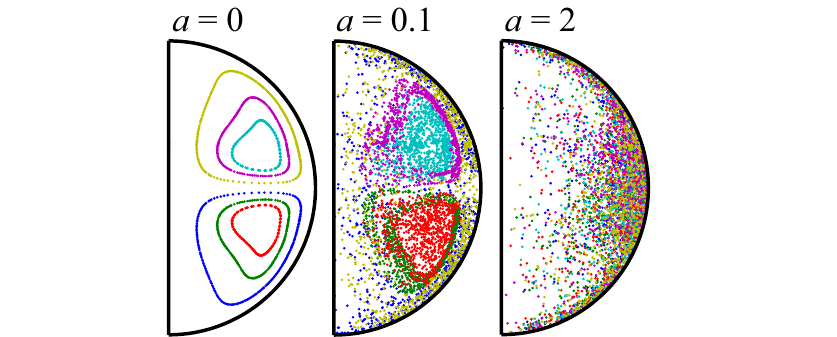}
\caption{Poincar\'e plots for the equilibrium flow $\v$ for $Re=300$, $\omega=0.6$ and three  amplitudes $a$. Points are the footprints of trajectories of six fluid parcels shown in different colors at one meridional plane. Each trajectory is followed for $1000T$.}\label{poincare}
\end{figure}

\begin{figure}[tbp]
\centering
\includegraphics[scale=1]{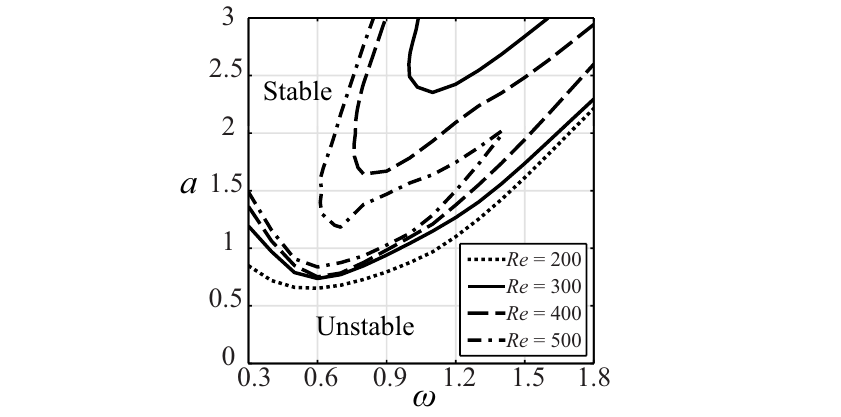}
\caption{Hydrodynamic stability boundaries of the equilibrium flow $\v$ on the plane of driving parameters $(\omega, a)$ for different fluid Reynolds numbers $Re$. For every $Re$, curves are composed of several azimuthal modes $m$ (typically, $m=1-4$). The shown region of parameters is stable for $Re<120$.}\label{HD_stab}
\end{figure}

An example of the calculated equilibrium velocity is shown in Fig.~\ref{tot_new2}(a).  The flow is axisymmetric  with well defined laminar structure at every instant of  time. However,  the paths of fluid elements in such flow are chaotic (Fig.~\ref{poincare}). From a practical point of view, it is important to have a hydrodynamically stable flow, so the flow  properties do not change over the course of experiment and  the predicted fast dynamos can be obtained. We study the linear stability of the flows by solving Eq.~(\ref{mhd_v}).  The stability boundaries  in the parameter space of $(\omega, a, Re)$ are shown in Fig.~\ref{HD_stab}. They are determined by non-axisymmetric modes only ($m=1-4$), the axisymmetric modes ($m=0$) are always stable for the considered flows. The stable region becomes smaller as fluid Reynolds number  $Re$ increases. We note that the steady counter-rotating flow corresponding to the boundary drive with  $a=0$  is linearly  unstable when  $Re>120$.  The time-periodic dependence of the flow plays twofold role: first, it stabilizes the fluid motion, and second, it leads to the chaotic streamlines required for the fast dynamo mechanism. The former is analogous to parametric stabilization observed in Kapitza's pendulum -- an inverted pendulum with pivot point vibrating in vertical direction \cite{Kapitza_1951}.

\begin{figure}[tbp]
\centering
\includegraphics[scale=1]{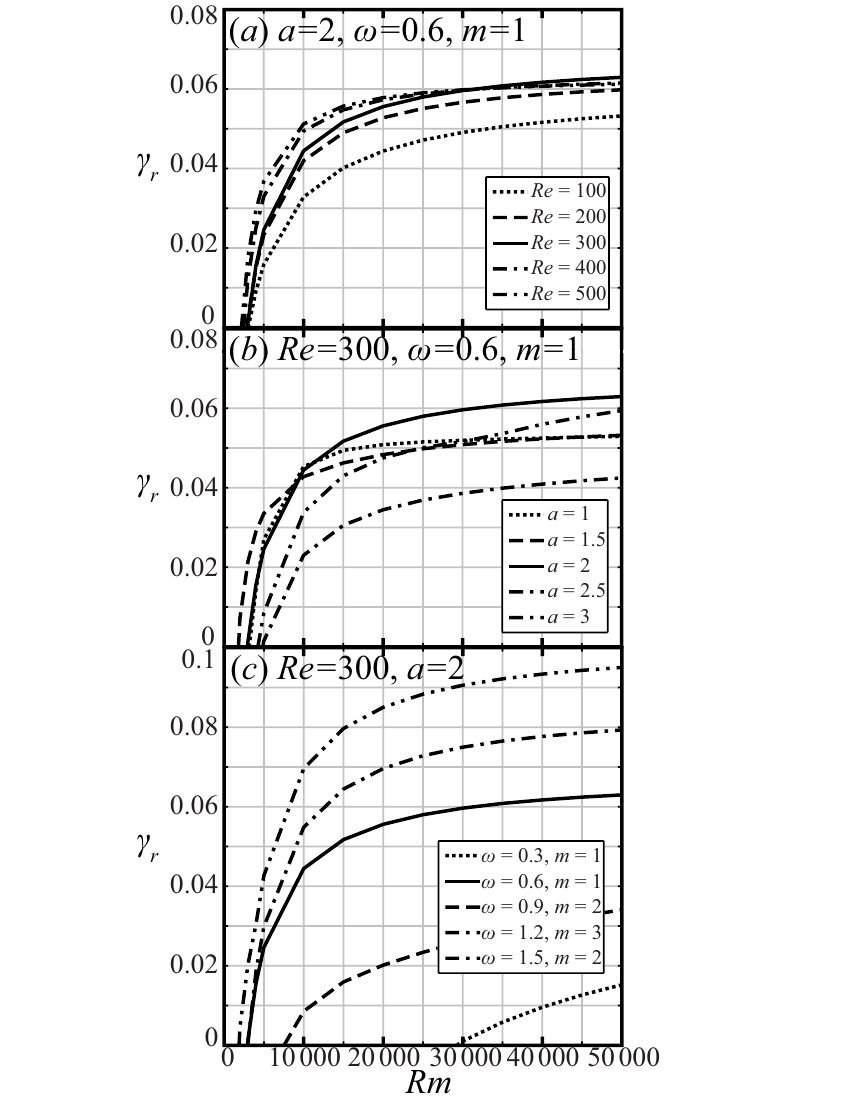}
\caption{Dynamo growth rate $\gamma_r$ of the fastest mode as a function of $Rm$ for different values of (a) fluid Reynolds number $Re$, (b) oscillation amplitude $a$, (c) driving frequency $\omega$. Note that in (a) and (b) the fastest azimuthal modes are always $m=1$, while in (c) they vary with $\omega$.}\label{gam_dyn}
\end{figure}

Fig.~\ref{gam_dyn} summarizes the results of our kinematic dynamo studies. The figure shows the dependences of the dynamo growth rate $\gamma_r$ (real part of the eigenvalue) on magnetic Reynolds number $Rm$ for different flow parameters. It is apparent  that in most cases the dependences tend to level off with increasing $Rm$. Such asymptotic behavior is a signature of fast dynamo action. Although none of these cases reaches its asymptotic state completely, they still suggest that fast dynamos are excited in the system. We note that the growth rates do not saturate until $Rm\sim10^5$, whereas in the Galloway-Proctor flow \cite{Galloway_1992} this occurs at $Rm\sim10^2-10^3$. A possible explanation was proposed  in Ref.~\cite{Hollerbach_1995}: in the spherical geometry the azimuthal number $m$ is restricted to integer values only, but in the plane-periodic geometry the corresponding wave-number is continuous  and can be optimized to yield the largest growth rate.  This also may explain why the typical critical $Rm$ required for the dynamo onset in our model ($Rm_{cr}\sim10^3$) is much higher than the one in the Galloway-Proctor flow ($Rm_{cr}\sim1$). Note that the value of  $Rm_{cr}\sim10^3$ can be achieved in MPDX by producing plasma with the electron temperature $T_e\sim10$~eV and  the driving velocity  $V_0\sim10$~km/s. 

The asymptotic values of the dynamo growth rates are determined by the flow properties. Most clearly this can be seen by changing driving frequency $\omega$  in Fig.~\ref{gam_dyn}(c). The largest value of $\gamma_r\approx0.1$ (in units of inverse turnover time $V_0/R_0$) is achieved with the largest frequency $\omega=1.5$. Simulations show that for $\omega\gtrsim1.7$ (at $Re=300, a=2$) the flow becomes hydrodynamically unstable, and our assumption of flow axisymmetry breaks.  Another interesting point is that the fastest dynamo modes correspond to different  azimuthal numbers $m$ ($m=1-3$) at different driving frequencies $\omega$. This is likely related to details of the equilibrium flow structure: as shown in Ref.~\cite{Richardson_2012}, the optimum $m$ depends on the number of poloidal flow cells,  which in our case varies with $\omega$.

\begin{figure}[tbp]
\centering
\includegraphics[scale=1]{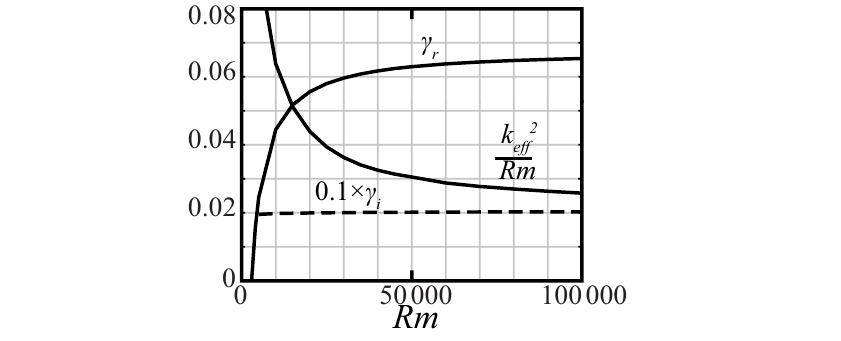}
\caption{Square of the effective wave-number  in a dynamo eigenmode $k_{eff}^2$ normalized by  $Rm$ and corresponding dynamo eigenvalue  $\gamma=\gamma_r+i\gamma_i$ as functions of $Rm$. Flow parameters are $Re=300$, $a=2$, $\omega=0.6$.}\label{k_eff}
\end{figure}

The dynamo eigenmode is shown in Fig.~\ref{tot_new2}(b). Due to our choice of the boundary conditions the magnetic field is fully contained inside the sphere. This is usually referred to as \emph{hidden} dynamo, because it  does not exhibit itself outside the bounded volume. The induced magnetic field has a fast-varying, small-scale  structure. Such field is often characterized by the magnetic Taylor microscale, defined as the inverse of  the effective wave-number $k_{eff}=(\langle |\nabla\times\B|^2\rangle/\langle|\B|^2\rangle)^{1/2}$, where $\langle\rangle$ denotes volume averaging. Fig.~\ref{k_eff} shows that for large $Rm$ the scaling law for $k_{eff}$ is approaching $k_{eff}\sim Rm^{1/2}$. Similar scaling is also observed in some slow dynamos \cite{Perkins_1987}. The difference between the slow dynamo as in Ref.~\cite{Perkins_1987} and the fast dynamo under consideration  is that in the former case the magnetic structures with spatial scales of the order of  $Rm^{-1/2}$  occur only in a small region near the separatrix of the flow, whereas in the latter case these structures are seen over a much larger region.    
  
In conclusion, we numerically demonstrated  the possibility of fast dynamo action in spherical boundary-driven time-periodic flows. This is an important step towards realizing the fast dynamos for the first time in plasma experiments such as MPDX.

The authors wish to thank  S.~Boldyrev, B.~P.~Brown, F.~Cattaneo and E.~Zweibel for valuable discussions. This work is supported by the NSF Award PHY 0821899 PFC Center for Magnetic Self Organization in Laboratory and Astrophysical Plasmas.

\end{document}